\documentclass[twocolumn,showpacs,preprintnumbers,amsmath,amssymb,prb,aps]{revtex4-1}

\usepackage{epsfig}% Include figure files
\usepackage{epstopdf}
\usepackage{multirow}
\usepackage{graphicx}% Include figure files
\usepackage{dcolumn}% Align table columns on decimal point
\usepackage{bm}% bold math
\usepackage{textcomp}
\usepackage{float}
\usepackage{array}
\usepackage{csquotes}
\usepackage{subcaption}
\usepackage{booktabs}
\usepackage{amsmath}
\usepackage{hyperref}% add hypertext capabilities
\hypersetup{
    colorlinks=true,
    linkcolor=blue,
    filecolor=magenta,      
    urlcolor=blue,
    pdftitle={},
    }
    
\begin{document}
\title{ Observation of correlation induced metal to half-metal phase transition and large orbital moment in $\mathrm{Sr_2CoO_4}$ }

\author{Shivani Bhardwaj$^{1,}$}
\altaffiliation{Electronic mail: spacetimeuniverse369@gmail.com}

\author{Sudhir K. Pandey$^{2,}$}
\altaffiliation{Electronic mail: sudhir@iitmandi.ac.in}
\affiliation{$^{1}$School of Physical Sciences, Indian Institute of Technology Mandi, Kamand - 175075, India}

\affiliation{$^{2}$School of Mechanical and Materials Engineering, Indian Institute of Technology Mandi, Kamand - 175075, India}

\date{\today} 

\begin{abstract}

We present a detailed mean-field study to address the fundamental discrepancy in the ground state magnetization of $\mathrm{Sr_{2}CoO_{4}}$ (SCO). In contrast to the ferromagnetic metallic ground state obtained from density functional theory (DFT), DFT+$U$ gives three ferromagnetic solutions converging to integer moment values (1, 2 \& 3 $\mu_B$/f.u) over a range of $U$. 
Interestingly, two of the solutions are found to exhibit half-metallicity with correspondingly $S$=1/2 and S=3/2 spin states. The half-metallic ferromagnetic state with $S$=3/2 is found to be the ground state solution for SCO. Co atoms show a large deviation from the formal +4 oxidation state indicating the presence of strong covalency effects.  
Our results suggest a plausible metal to half-metal phase transition around $U$($J$)=4.4(1.16) eV. The Fermi surface study shows gradual collapse in states leading to half-metallicity suggesting  \textit{\textbf{k}}-dependence of effective $U$ around the critical region. Surprisingly, in the presence of spin-orbit coupling (SOC), unexpectedly large orbital moment ($L_{z}$=0.6) is noted in SCO putting it among the class of 3$d$ based transition metal compounds exhibiting pronounced orbital magnetization. The calculations give large magnetocrystalline anisotropy energy (MAE) of $\sim$48 meV. 
Large values of orbital magnetic moment contribution and MAE, in the presence of strong correlation effects, provide a better interpretation of experimental magnetization observed in SCO. 

\vspace{0.2cm}

\end{abstract}

\maketitle

\section{Introduction}
 Layered pervoskites especially with structures akin to $\mathrm{K_{2}NiF_{4}}$ \cite{0} have been extensively studied after the discovery of superconductivity in $\mathrm{La_{2-{x}}Sr_{x}CuO_{4}}$\cite{Imada} and $\mathrm{Sr_{2}RuO_{4}}$ \cite{Maeno}, for showcasing intriguing properties ranging from spin and/or orbital order\cite{1986,11,f,2005,2006_1,2006_2,srvoo4,2020}, spin/charge stripes formation in nickelates \& manganites\cite{Imada}, and spin-triplet superconductivity in ruthenates \cite{Maeno}.
 The discovery of superconductivity in layered Cu oxides with two dimensional $\mathrm{CuO_{2}}$ sheets\cite{2000} tempted researchers to investigate several low dimensional $\mathrm{CoO_{2}}$ structures. $\mathrm{Sr_{2}CoO_{4}}$ (SCO), a quasi two-dimensional layered perovskite, is reported as the only member of  $\mathrm{K_{2}NiF_{4}}$ family to show ferromagnetic spin-response and metallic behavior\cite{2004prl}.

Earlier experimental studies on SCO (formally $\mathrm{Co^{4+}}$) reported contradictory magnetization states having 1.8 $\mu_B$/$Co$\cite{2004prl} and 1 $\mu_B$/$Co$\cite{2005w2} saturation magnetization values with high Curie temperature (250 K).
Notably, $Wang$ $et$ $al$'s experimental study\cite{2005w2} also shows a discrepancy between the saturation magnetization value obtained from the $M(H)$ curve and that from the effective magnetic moment (MM), ($p_{eff}$=3.77 $\mu_B$, $S$=3/2). 
The difference in observed magnetization in the aforementioned works sparked interest in pursuing the ground state magnetization inherent to SCO\cite{}. Subsequent theoretical studies predict half-metallicity although with contrasting spin-states\cite{2006,2010,2017_1}. A recent experimental study shows presence of stair-case like magnetization behavior along with high coercive field suggesting the possibility of multiple magnetic phases and presence of large MAE in SCO\cite{2016}. In later experimental study by $Pandey$ $et$ $al$, saturation magnetization even lower than 1 $\mu_B$/$Co$ was reported with rather higher $p_{eff}$=4.26 $\mu_B$, claiming presence of $\mathrm{Co^{4+}}$ as mixture of intermediate and high spin state\cite{2013_p2}. Moreover, we see the difference in saturation magnetization values in the experimental studies  primarily due to the lack of true saturation of $M(H)$ curves \cite{2004prl, 2005w2,2013_p2} within the specified range of external magnetic fields.\\
$Lee$ $et$ $al$\cite{2006} showed a saturation magnetization of approximately 2 $\mu_B$/f.u using both LDA and LDA+$U$ methods. However, their LDA+$U$ calculation suggests the presence of a half-metallic state with MM of 1 $\mu_B$/f.u beyond a critical value of $U$=2.5 eV. Their work proposes a HM solution with band gap of $\sim$0.25 eV in the majority $e_{g}$ sector. In another mean-field study by GGA+$U$ over a range of $U$ with fixed $J$ (0.4 eV), a ferromagnetic HM ground state with $S=3/2$ is reported with a large gap ($\sim$1 eV) in the minority $t_{2g}$ sector arising out of competing antiferromagnetic and ferromagnetic Co ion interactions \cite{2010}. The studies suggest importance of correlation effects in determining the electronic and magnetic properties of SCO. Apart from the difference in exchange-correlation functional used i.e. LDA \& GGA, the above two studies also differ in the value of $J$ employed over the range of $U$. The magnetization solution of $S$=3/2 is however reported to exist regardless of the type of exchange-correlation functional used\cite{2010}. Thus, the different magnetization solutions achieved in these studies might be attributed to the $J$ parameter used.
Also from the literature, magnetic properties in strongly correlated electron systems are shown to be particularly sensitive to the $J$ value; thus, proper choice of $J$ becomes crucial when interpreting magnetization in such system\cite{J,antik}. 
Furthermore, in DFT+$U$ formulation, various local minima are accessible corresponding to potential solution states and hence attempting DFT+$U$ requires careful identification of the converged solution to be considered for global minima/ground state solution \cite{1,2,3,4}. Consequently, the disparity in proposed DFT+$U$ solutions available on SCO could be ascribed to the neglect of above mentioned features of DFT+$U$ methodology. It is also worth mentioning here that a half metallic solution has been recently obtained in SCO using mBJ exchange potential with a relatively high band gap of 1 eV in the minority states, yielding a spin-MM of 1.6 $\mu_B$/Co\cite{2017}. \\
Conclusively, there exists variation in magnetization solutions in the theoretical studies where there is large difference between the calculated spin-MMs and the spin moment obtained from the experimental $p_{eff}$. For instance, considering a spin moment of 3 $\mu_B$/Co or higher as indicated by the $p_{eff}$ value in experimental magnetization reports\cite{2005w2,2013_p2} contradicts the observed saturation magnetization value which is far lower than 3 $\mu_B$/f.u. Typically such discrepancy in interpreting the total magnetization of the system is bridged by considering orbital moment contribution\cite{Marshall,Terakaru,nio,Huang,Lee_1,Suzuki,Tulika,lal,l,s,r}. Moreover, numerous recent studies of last decade have shown Co atoms exhibiting large orbital magnetization in several Co based layered oxides\cite{q,p,c,b,z,2019}. Therefore, it will be interesting to study the role of orbital degrees of freedom in understanding the true magentization of SCO.\\
Accordingly, in this work we report electronic structure calculations of SCO via a comprehensive mean-field study incorporating correlation effects and SOC. Calculations in both the non-magnetic (NM) and Ferromgnetic (FM) phases are carried out within DFT framework. Subsequently, the role of electronic correlations in the system is systematically studied using DFT+$U$ in the varying $U$ (correspondingly $J$) regime along with Fermi surface (FS) description. Further section of the work is devoted to studying the significance of SOC in understanding the magnetic properties of SCO. Our work provides insight into studying the transition metal complexes considering the participation of orbital degrees of freedom which are otherwise considered insignificant in 3$d$ based transition metal compounds. \\
\section{COMPUTATIONAL DETAILS }

The electronic structure calculations of SCO within DFT and DFT+$U$ frameworks are carried out with PBEsol exchange functional\cite{pbesol} using elk code\cite{elk}. An all-electron full-potential linearised augmented-plane wave method is used for the calculations. In the calculations, muffin-tin radii of 2.25, 1.82 and 1.57 a.u. are used for Sr, Co and O atoms, respectively. Fully localized limit- double counting scheme is used in the calculations\cite{anni}. The Brillouin zone is sampled by a regular mesh containing up to
204 irreducible k points for the calculations. The energy convergence criterion used is $\mathrm{10^{-4}}$ Ry/cell. Methfessel-Paxton order 2 - type of smearing approach is used for smooth approximation to the Dirac delta function for computing the occupancies of the Kohn-Sham states as implemented in the elk code. The SOC is further employed on DFT+$U$ results to study the role of orbital degrees of freedom. The FS depiction is plotted using XCrysDen\cite{xcrysden}.

\section{Results and Discussion}

\begin{table}[h]
    \centering 
    \caption{\small (a) Relaxed atomic positions of Sr and O2 in NM and FM phases (b)  Bond-distance (in $A^{o}$) between Co-O atoms in NM and FM phases of $\mathrm{Sr_{2}CoO_{4}}$.}
     
     \begin{tabular}{|c|c|c|c|c|}
    
      \hline
      \tiny (a) Atom & \tiny NM & \tiny FM   \\
     \hline
      \tiny Sr & \tiny 0.6423 & \tiny 0.64321 \\
     
      \tiny O2 & \tiny 0.8445 & \tiny 0.8436 \\
        \hline
         \end{tabular}
     \centering 
    \begin{tabular}{|c|c|c|c|c|}
   
      \hline
      \tiny (b) Bond distance & \tiny NM & \tiny FM   \\
       \hline
      \tiny Co-O1 & \tiny 1.898 & \tiny 1.898  \\
        
      \tiny Co-O2 & \tiny 1.941 & \tiny 1.953 \\
       \hline
    
     \end{tabular}
    
    \label{a}
\end{table}
\vspace{1.0cm}

$\mathrm{Sr_{2}CoO_{4}}$ crystallizes in $bcc$ tetragonal structure belonging to space group number $139$ (\textit{I4/mmm)}. It consists of Sr atoms with variable positions at $4e$ (0,0,z), two inequivalent O atoms- O1 (planar) and O2 (apical) at $2b$ (0.5,0,0) and $4e$ (0,0,z), respectively wherein Co atoms occupy 2a (0,0,0) wyckoff positions. Relaxed positions of Sr and O2 in NM (FM) phase are estimated to be 0.6423 (0.64321) and 0.8445 (0.8436) using the experimental lattice parameters (a=b=3.79 $A^{0}$, c=12.48 $A^{0}$)\cite{2005w2} in both the NM and FM phases.
The relaxed positions of Sr and O2 used further in calculations and estimates of Co-O1 and Co-O2 bond lengths obtained are provided in Table I.
As evident from the Table I, Co-O2 (1.953 $A^{o}$) bond length in FM phase is found to be  0.6\% larger than that in the NM phase (1.941 $A^{o}$). The calculated Co-O1 \& Co-O2 bond lengths are found to be consistent with the experimentally deduced bond-lengths\cite{2005w2}.
The unequal bond lengths of Co-O1 and Co-O2 suggest distorted octahedral geometry of $\mathrm{CoO_{6}}$ in FM phase, and can be considered due to ferrodistortive Jahn-teller ordering as mentioned in earlier studies\cite{2010}. 

In order to study the ionic states, crystal field splitting and nature of bonding, partial density of states (PDOS) of Sr (5s), Co (3$d$,4$s$) and O (2$p$) atoms have been calculated in NM phase and given in Fig. 1. The number of electrons in Sr 5s state has been found to be negligible (0.05), indicating its as expected +2 oxidation state.
From the Table II, Co 3$d$ orbitals contains a total of $\sim$6.25 number of electrons, thus Co atoms can be regarded to be present in less than +3 oxidation state. The contribution of O1 and O2 (2$p$) states is approximately 3.94 and 3.88, respectively, while the available states for O1 (O2) 2$p$ in the unoccupied band come out to be $\sim 0.49$ ($\sim 0.48$). Therefore, both the O1 and O2 must be present in less than -2 oxidation states. The difference in number of electrons corresponding to O1 2$p$ ($\sim 3.438$) and O2 2$p$ ($\sim 3.393$) (see Table II) traces back to the difference in Co-O1(O2) bond lengths (see Table I).
Based on the criterion that maximum number of states should be occupied in lowest energy states, the highest orbital occupancy of Co $d_{xy}$ (1.563) indicates it to be the lowest energy state, followed by next lower occupancy of $d_{yz}/ d_{xz}$ (1.547) revealing the splitting of $t_{2g}$ orbitals into $d_{xy}$ and $d_{yz}/d_{zx}$. Subsequently, according to the occupancy of $d_{z^2}$ (0.803) and $d_{x^{2}-y^{2}}$ (0.79), $e_{g}$ sector is expected to be present in higher energy region. 
However, the location of orbitals in energy based on their occupancy order is found to be inconsistent with their location observed in the PDOS. 
For instance from Fig. 1, the $e_{g}$ states ($d_{z^{2}}$ and $d_{x^{2}-y^{2}}$) spread to lower energies even below the extent of $t_{2g}$ states in PDOS. The low lying Co $e_{g}$ states indicate the possibility of large extent of overlap of $e_{g}$ orbitals with 2$p$ orbitals of O atoms. The appearance of large energy gap between occupied and unoccupied $e_g$ states in the PDOS also indicates strong hybridization with O 2$p$ orbitals.
 Based on the oxidation states of Co and O atoms, occupancy and position of Co 3$d$ orbitals, it can be remarked that here purely ionic model would fail and strong co-valency in Co and O bonds is expected to play major role in deciding the electronic and magnetic properties of this compound.

 \begin{table}[h]
 \addtolength{\tabcolsep}{0.04pt}
    \centering 
  
    \caption{\small Orbital (Orb) occupancy (Occ) of Co $d_{xy}$, $d_{yz}, d_{zx}$ , $d_{x^{2}-y^{2}}$, $d_{z^{2}}$ orbitals and  $p_x$, $p_y$, $p_z$ orbitals of O1 \& O2  in NM phase.}

    \begin{tabular}{|c|c|c|c|c|c|c|c|c|c|}

      \hline
      \tiny Orb &  \tiny $d_{xy}$ & \tiny $d_{yz}/ d_{xz}$  & \tiny $d_{x^{2}-y^{2}}$ & \tiny $d_{z^{2}}$ & \tiny O1 ${p_{x}}$ & \tiny O1 ${p_{y}}$ & \tiny O1 ${p_{z}}$ & \tiny O2 ${p_{z}}$ & \tiny O2 ${p_{x}/p_{y}}$ \\
      \hline
 \tiny Occ & \tiny 1.563  & \tiny 1.547 & \tiny 0.79 & \tiny 0.803 & \tiny 1.227 & \tiny 1.00 & \tiny 1.211 & \tiny 1.018  & \tiny 1.187   \\
%\tiny Occupancy & \tiny 1.563  & \tiny 3.095 & \tiny 0.79 & \tiny 0.803 & \tiny 1.227(0.0685) & \tiny 1.00(0.32964) & \tiny 1.211 (0.09858)& \tiny 1.018 (0.27689) & \tiny 2.375(0.21204)    \\  
        \hline
     \end{tabular}
        
    \label{tab:}
\end{table}
\vspace{1.0cm}
\begin{figure}
    \centering
    \includegraphics[width=6cm]{nm-dos.eps}
    
    \caption{ \small DFT calculated PDOS of Co 3$d$ and O1 and O2 2$p$, in NM phase.}.

    \label{fig:}
\end{figure}

 Since, the experiments report FM state for SCO, we have carried out calculations in FM phase to study its spin-resolved distribution of states and exchange splitting energy.   
The total energy corresponding to FM structure is found to be lower than NM structure by 0.32 eV/f.u., suggesting it to be ground state of the system. The PDOS calculated for the Co 3$d$ and O2 2$p$ orbitals for both the majority and minority channels is given in Fig. 5 given in supplementary. We see that similar to the NM phase, here also the $e_{g}$ states spread to lower energy region in both the majority and minority channels suggesting the consistency with aforementioned scenario. We find that shift in the edge of total density of states in FM case w.r.t NM case around their respective Fermi levels (on the absolute scale of energy) could be a rough estimate of effective Hund's like exchange splitting and is found to be $\sim$0.53 eV. The total spin-MM is found to be $\sim$2 $\mu_B$/f.u. Total MM majorly comes from Co atoms ($\sim$1.56 $\mu_{B}$/Co) as expected. Notably, O1 is also found to contain substantial MM ($\sim$0.16 $\mu_{B}$/atom) indicating strong hybridization effects. The PDOS calculations show that the vicinity of Fermi-level is mostly comprised of Co $d_{xz}/d_{yz}$ and $d_{xy}$ minority states whereas, negligible contribution comes from the majority channel thus indicating its almost half-metallic state.

\begin{figure}
    \centering
    \includegraphics[width=8cm]{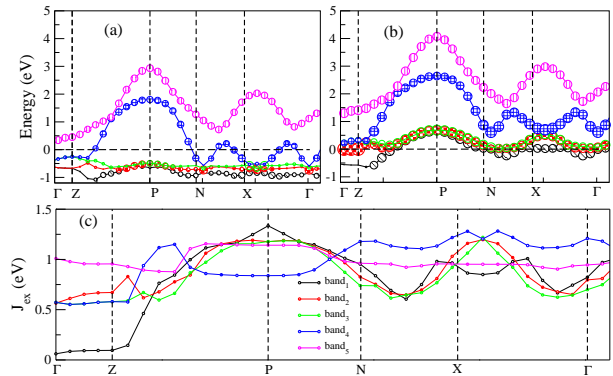}
    \caption{\small Band structure of UP (a) and DN (b) states, where the circle size represents the weight of 3$d$ character in the bands. (c) Exchange splitting energy ($J_{ex}$) along the high-symmetric k-directions. }.

    \label{fig:}
\end{figure}
Further, exchange splitting energy of the Co 3$d$ orbitals, along high symmetric k-directions for bands around Fermi-level is plotted in Fig. 3. The energy difference between the UP and DN band, gives rough idea of exchange splitting energy ($J_{ex}$). Here, the highest value of $J_{ex}$(1.32 eV) is found around P point corresponding to $band_{1}$. We see that, $band_{4}$ shows high exchange splitting along X to N path, giving higher $J_{ex}$ at N (1.17 eV) and X (1.24 eV) points. As evident from the figure, throughout the path along high symmetric k-directions, the $J_{ex}$ value is consistently above 0.5 eV except for $band_{1}$, which gives value less than 0.5 eV at $\Gamma$ and Z points. Evidently, $band_1$ has the lowest of all exchange splitting energy at Z point (0.11 eV). Around P point $band_{1}$, $band_{2}$ \& $band_{3}$ have relatively equal weight of 3$d$ character in their respective UP and DN bands, thus the energy difference of their respective UP and DN bands is near about same and hence the value of $J_{ex}$. Conclusively, the estimate of $J_{ex}$ could be reasoned on the basis of weight of 3$d$ character present in the bands.

Although DFT in FM phase is able to provide the observed metallic ferromagnetic state of this system\cite{2004prl,2005w2} however, DFT is known to fail for the proper description of strongly correlated systems. Thus, we have carried out DFT+$U$ calculations to study the realistic estimates of its electronic structure properties and the role of electronic correlations in SCO.

\begin{table*}[htbp!]
    \small\addtolength{\tabcolsep}{5pt}
    \caption{ \small (a)   Orbital occupancy of Co 3$d$ orbitals $d_{xy}$, $d_{yz}/d_{zx}$ , $d_{x^{2}-y^{2}}$, $d_{z^{2}}$ for UP (DN) channels obtained at DFT+$U$ level for $U$=5 eV. (b) MM values ($\mu_B/atom$) for atoms and magnetization of interstitial \& total value ($\mu_B$/f.u) for S1, S2 and S3 solutions at $U$=5 eV.}
    \begin{tabular}{|c|c|}
    \hline
      \tiny (a) Orbital occupancy UP (DN) \\
    
    \hline
    \begin{tabular}{c|c|c|c|c|c} 
    \small\addtolength{\tabcolsep}{5pt}
     \tiny Solutions & \tiny $d_{xy}$ & \tiny $d_{yz}$/$d_{xz}$  & \tiny $d_{x^{2}-y^{2}}$  & \tiny $d_{z^{2}}$  & \tiny total  \\
      \hline
      \tiny S1 & \tiny 0.907 (0.072) & \tiny 0.897 (0.853)  & \tiny 0.262 (0.629) & \tiny 0.350 (0.389)  & \tiny 3.313 (2.796) \\
      \tiny S2 & \tiny 0.909 (0.261)   & \tiny 0.904 (0.758)  & \tiny 0.773 (0.152) & \tiny 0.552 (0.276) & \tiny  4.042 (2.205)   \\
      \tiny S3 & \tiny 0.911 (0.859) & \tiny 0.912 (0.245) & \tiny 0.753 (0.210) & \tiny 0.787 (0.361)  & \tiny  4.275 (1.920) \\
  
        \hline
   \end{tabular}
         \end{tabular}  
    \label{tab:}
\end{table*}

 \begin{figure}
    \centering
    \includegraphics[width=9cm]{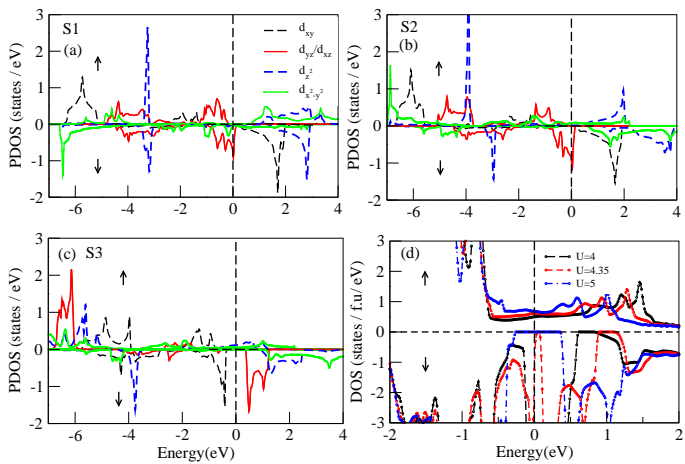}
    \caption{\small DFT+$U$ calculated FM PDOS of Co 3$d$ orbitals at $U$= 5 eV for (a) S1 (b) S2 and (c) S3 solutions. (d) Comparative behavior of Co 3$d$ total DOS in the vicinity of Fermi-level at $U$=4, 4.35 and 5 eV.}

    \label{fig:}
\end{figure}

In DFT+$U$ calculations, $U$ and $J$ are used as parameters, and determination of accurate value of these parameters for a given material is still theoretically challenging.
Additionally, the choice of $U$ and $J$ values are found to depend on the approximations employed in the implementation\cite{pandey2008,blaha}. 
 We have thus studied the system over a range of $U$ values (0-6 eV), where the corresponding $J$ is calculated through Yukawa screening method\cite{yuk,anni}. 
In DFT+$U$ method, the calculations are often found to converge in different local minima depending on different starting electronic distribution, which indicates its dependence on the starting point (electronic configuration) \cite{1,2,3}. Thus, the solution obtained from single starting electronic distribution in the correlated orbital may not correspond to the ground state solution. In this situation, to find ground state of the system, one needs to find various possible solutions existing for the system. This can be done through different starting electronic distribution in the 3$d$ orbitals.\\

\begin{table*}[htbp!]
  \small\addtolength{\tabcolsep}{5pt}
    \centering

     \begin{tabular}{|c|c|c|c|c|c|c|c|c|c|}
      \hline
      \tiny (b)  Solutions &  \tiny Co  & \tiny O1   & \tiny O2 & \tiny Interstitial & \tiny total  \\
      \hline
  \tiny S1 & \tiny 0.330 & \tiny 0.193  & \tiny 0.097  & \tiny 0.138 & \tiny  1.054 \\
\tiny S2 & \tiny 1.772  & \tiny 0.108  & \tiny -0.049 & \tiny 0.114 & \tiny  2.008  \\
  \tiny S3 & \tiny  2.365 & \tiny 0.076 & \tiny 0.125 & \tiny 0.228 & \tiny  3.008 \\

        \hline
     \end{tabular}
    \label{tab:}
\end{table*} 
Thus, we have strategically tried to find all the possible solutions existing for SCO starting with different possible electronic configurations. We find three different solutions which are categorized into S1, S2 and S3 based on the magnetization value and the density matrix (occupancy distribution of 3$d$ states). 
While, S1 and S3 solutions vanish below $U$=3.8 eV and $U$=2 eV, respectively. All the three solutions are found to exist simultaneously above $U$=3.8 eV. However, S2 solution is found to persist throughout the studied $U$ range. S1 and S2 solutions correspond to magnetization values of $\sim$1 $\mu_B$/f.u and $\sim$2 $\mu_B$/f.u, respectively. The magnetization values of both the solutions remain robust with varying $U$. Unlike S1 and S2, S3 solution is found to exhibit varying magnetization over the range of $U$. The magnetization shows continuous increment in value from $\tiny{\sim}$2.5 $\mu_B$/f.u to $\sim$3.2 $\mu_B$/f.u with increasing $U$(2-6 eV) in case of S3 solution. Across the $U$ range, S2 solution is metallic and S1 solution is found to be HM. Whereas, S3 solution exhibits half-metallicity above a critical value of $U$=$\sim$4.4 eV. Notably, among these three solutions, S3 is found to have minimum energy throughout the studied $U$ range, indicating it to be the ground state of the system. For the detailed depiction of nature of S1, S2 and S3 solutions, magnetization values and occupancy of Co 3$d$ states are provided in Table III along with PDOS given in Fig. 3. 
\\
\begin{table*}[hbtp!]
  \small\addtolength{\tabcolsep}{7pt}
    \centering 
    \caption{\small Total magnetic moment values ($\mu_B$/f.u), atomic MM of Co, O1 \& O2 atoms ($\mu_B$/atom) and phase information (M for metallic and HM for half-metallic), for S3 with $U$ variation (0-6 eV). }
  
     \begin{tabular}{|c|c|c|c|c|c|c|c|c|c|c|}
      \hline
      \tiny $U$ &  \tiny 0  & \tiny 2   & \tiny 3 & \tiny 4 & \tiny 4.3 &\tiny 4.4 &  \tiny 4.5  & \tiny 5 & \tiny 6  \\
      \hline

\tiny Co & \tiny 1.55 & \tiny 1.74 & \tiny 1.88 & \tiny 1.97  & \tiny 2.10 & \tiny 2.14 & \tiny 2.18  & \tiny 2.36 & \tiny 2.60 \\

\tiny O1 & \tiny 0.16 & \tiny 0.14 & \tiny 0.15 & \tiny  0.14 & \tiny  0.11&   \tiny 0.09 &  \tiny 0.09  &\tiny 0.07 & \tiny 0.06 \\

\tiny O2 & \tiny 0.003 &  \tiny 0.07 & \tiny 0.07 & \tiny  0.07 & \tiny 0.09  &  \tiny 0.10 &  \tiny 0.11 & \tiny 0.12 & \tiny 0.15 \\

%\tiny Interstitial &  \tiny 0.10 & \tiny 0.20 & \tiny  0.20 & \tiny  0.20  & \tiny 0.21 & \tiny 0.21  & \tiny 0.21  & \tiny 0.22 & \tiny 0.24\\ 

 \tiny Total MM  & \tiny 2.00 & \tiny 2.37 & \tiny 2.55  & \tiny 2.61 & \tiny 2.73 & \tiny 2.77 & \tiny 2.81 & \tiny 3.00 & \tiny 3.27 \\
  \hline
 \tiny Phase & \tiny M  & \tiny M  & \tiny M & \tiny M & \tiny  M & \tiny HM & \tiny HM & \tiny HM & \tiny HM \\

        \hline
     \end{tabular}  
    \label{tab:}
\end{table*}
As mentioned above, S2 is present throughout the varying $U$ regime and is the only solution that exists at the uncorrelated level ($U$=0 eV). However, S1 and S3 arise due to the inclusion of correlation effects in SCO.\\
It is quite intriguing to learn how $U$ leads to other potential solutions of SCO that are not otherwise achieved at the uncorrelated level.
The orbital occupancy data for each solution suggests that increasing $U$ does not change the total number of electrons in the 3$d$ orbitals, but introduces rearrangement in occupancy and hence the magnetic moments.
In order to understand the origin of S1 and S3, it would be important to study their fundamental characteristics with respect to S2.
Thus we have established comparison and characteristics of S1, S2 and S3 solutions at $U$=5 eV in the discussion below.\\
On comparing the Co 3$d$ orbitals' occupancy and MM on O atoms, S1 is observed to share similarity with S2 in certain respects.   
In both the solutions, degenerate $d_{xz}$ \& $d_{yz}$ orbitals contain highest number of electrons, as evident from Table III (a). Also, in both the cases, planar O atoms are found to carry higher MM than apical O atoms. In contrast to this, in S3, $d_{xy}$ orbital has the highest occupancy and apical O atoms have higher MM than planar O atoms.

For S1, the distribution of electrons in UP and DN states in different 3$d$ orbitals is near about same as that in S2 (qualitatively in terms of more than or less than half-filled) except for $d_{x^{2}-y^{2}}$. The number of electrons in DN state are larger than those in UP state of $d_{x^{2}-y^{2}}$, which is mainly responsible for lowered total magnetization of S1 than S2. This is also evident from remarkably reduced MM on Co atom ($\sim$0.33 $\mu_B$/Co). Thus, the reason for possibility of reduced magnetization for S1 (1 $\mu_B$/f.u) could be seen as the reversed occupancy behavior of $d_{x^{2}-y^{2}}$ orbital as compared to S2.
In case of S3, what makes it different from S2 is that the highest occupancy is no longer in degenerate orbitals ($d_{yz}$ \& $d_{xz}$). Additionally the DN channel in degenerate orbitals is less than half-filled which contributes in raising the magnetization. The presence of higher MM on Co atom in S3 ($\sim$2.36 $\mu_B$/f.u) proves the same. Thus, the changed occupancy behavior of DN channel in $d_{yz}$ and $d_{xz}$ results in increased total magnetization value (3 $\mu_B$/f.u) of S3  in comparison to S2. Additionally, the presence of negative spin-moment on the apical O atoms in S2, makes it different from the other two solutions.\\
From the Fig. 3, it is clear that S1 and S3 solutions are HM, with gap appearing in majority ($\sim$0.75 eV) and minority channels ($\sim$0.6 eV), respectively. Evidently, $d_{x^2-y^2}$ in S1 and $d_{yz}$ \& $d_{xz}$ in S3 are observed to participate in the gap formation. Conclusively, the orbitals previously held responsible for the changed magnetization values of S1 and S3 with respect to S2, contribute mainly in deciding the HM phase of these solutions.  
In terms of energy, S1 and S2 lie closely spaced differing by $\sim$180 meV. The energy comparison suggests S3 is the ground state, and S2 \& S1 are the first and second excited states, respectively for SCO.

As stated earlier, S3 is HM beyond the critical value of $U$=$\sim$4.4 eV, where the gap goes on increasing with increasing $U$ along with gradual increase in magnetization value ($\sim$3.3 $\mu_B$/f.u at $U$=6 eV). The HM S3 state is identified to be an intermediate spin state ($S$=3/2) based on its magnetization value ($\sim$3 $\mu_B$/f.u) and Co 3$d$ occupancy distribution (between UP and DN channels). Below $U$=4.4 eV, it is metallic and the magnetization value decreases with decreasing $U$, eventually reduces to $\sim$2.5 $\mu_B$/f.u at $U$=2 eV. A remarkable feature to note from Table IV, is the changed behavior of MM on O atoms at the critical point ($U$=4.4 eV). Below the critical value, planar O atoms show relatively higher MM than apical ones; whereas towards the critical point the moment gradually starts to shift on the apical O atoms and becomes near about equal to planar ones. Interestingly, above the critical $U$, the MM appears to be substantially pushed on apical O atoms from the planar O atoms. We find this striking crossover in the MM value from planar to apical O atoms, as the key feature responsible for metallic to HM phase transition.\\

On comparing with previous studies, we find our HM S1 solution identical to HM solution reported earlier by $Lee$ $et$ $al$ $\cite{2006}$. The similarity can be particularly noted in terms of occupancy distribution in Co 3$d$ and magnetization value. The characteristic feature of negative contribution from $d_{x^2-y^2}$ orbital (DN channel more occupied than UP) along with diminishing contribution from $d_{z^2}$ in S1, is also found in their HM solution \cite{2006}.
In contrast to the HM solution reported by  $Lee$ $et$ $al$ which is claimed to undergo collapse in its magnetization with increasing $U$, S1 remains intact with its magnetization of 1 $\mu_B$/f.u with increasing $U$. Moreover, in their case coexistence of both metallic and HM solutions (which they point out as high-moment and low-moment solutions) is reported only at a critical $U$=2.5 eV, above which only HM solution with gradually collapsing moment is claimed. 
However, the decrease in MM on Co atom and increase in moment on planar O atoms with $U$ is noted in our S1 solution as well. Apparently, the presence of anti-parallel alignment of moments on planar O atoms in their case results in gradually collapsing magnetization with $U$, unlike our S1 where magnetization value remains robust with varying $U$. Importantly, we discover that both S1 and S2 can be stabilized beyond the critical value (3.9 eV) in the examined $U$ range, with the high moment state (S2) having lower energy than S1. However, this is not the case with $Lee$ $et$ $al$'s solutions.

Furthermore, the HM ground state solution (S3) reported in this work, resembles the HM ground state solution in the previous study by $Pandey$ \cite{2010}. Their HM solution likewise our S3 solution, is also characterized by less than half-occupancy of $d_{yz}$ \& $d_{xz}$ states together with nearly filled $d_{xy}$ state. Notably, our HM S3 solution is found to exhibit varying magnetization with $U$ and also becomes metallic below a critical value unlike their result where they do not find any such variation with $U$. The reason for their magnetization value (3 $\mu_B$) remaining intact with varying $U$ can be attributed to their calculations being carried out at fixed $J$ value (0.4 eV) across their investigated $U$ (3-5 eV) range. Whereas, in our case we have used the corresponding $J$ values for each $U$, calculated using Yukawa screening method as stated earlier in the text.  Evidently, from Table VI given in supplementary, significant change in $J$ value with $U$ explains the changing magnetization observed in our case (S3) over the studied $U$ range.\\

In one aspect the magnetization behavior of HM solutions (S1 \& S3) reported in this work, differs from the HM solutions obtained in aforementioned DFT+$U$ studies due to difference in application of $J$ over range of $U$. It would be unrealistic to carry out calculations at fixed $J$ for whole range of $U$ in a material like SCO, where $J$ value is observed to change significantly with $U$ ( refer Table VI given in supplementary). 
It is noteworthy to mention that utilizing characteristic features of DFT+$U$ methodology as discussed in the text, we could innovatively retrieve all the possible solution states to finally comment on the total magnetization and ground state spin configuration of SCO. 
Our study succeeds in combining the earlier mean-field studies, which were perceived to propose contradictory magnetization states. As a result we are able establish the true ground state by obtaining all possible solutions for SCO.

 \begin{table*}[htbp!]
  \small\addtolength{\tabcolsep}{5pt}
    \centering 
    \caption{ \small Variation of $S$, $L$, $J$, saturation magnetization $\mu_z$ ($\mu_B$/f.u), $p_{eff}$ ($\mu_B$) with $U$ (eV) along $\left\langle 001 \right\rangle$ direction .}
  
     \begin{tabular}{|c|c|c|c|c|c|c|c|c|c|}
      \hline
      \tiny $U$ &  \tiny 3  &\tiny 4 & \tiny 4.3   & \tiny 4.4 & \tiny 4.5 & \tiny 5  \\
      \hline
  \tiny $L$ & \tiny 0.21 & \tiny 0.53 & \tiny 0.61  & \tiny 0.55  & \tiny 0.53 & \tiny 0.48 \\
  \tiny $S$ & \tiny 0.93 & \tiny 0.98 & \tiny 0.97  & \tiny 0.88  & \tiny 0.83 & \tiny 0.77 \\
  \tiny $J$  & \tiny 1.15 & \tiny 1.52 & \tiny 1.58  & \tiny 1.43   & \tiny 1.37 & \tiny 1.25 \\
  \tiny $\mu_z$ & \tiny 2.08 & \tiny 2.51 & \tiny 2.56  & \tiny 2.31   & \tiny 2.20 & \tiny 2.02 \\
  \tiny $p_{eff}$ & \tiny 2.85 & \tiny 3.23  &  \tiny 3.27  & \tiny 3.0  & \tiny 2.9  & \tiny 2.71 \\
        \hline
     \end{tabular} 
   % \label{tab:}\\
  
\end{table*}

We have also examined the effect of SOC on the ground state solution $i.e.$ S3, along $\left\langle 001 \right\rangle$, $\left\langle 100 \right\rangle$, $\left\langle 101 \right\rangle$, $\left\langle 110 \right\rangle$, $\left\langle 111 \right\rangle$ directions. The energy of $\left\langle 001 \right\rangle$ direction is found to be lowest and $\left\langle 111 \right\rangle$ direction having the next lowest energy \& $\left\langle 100 \right\rangle$ the third lowest for all values of $U$ except at $U$= 5 eV. At $U$=5 eV, $\left\langle 111 \right\rangle$ direction is found to have lowest energy with $\left\langle 001 \right\rangle$ direction having next lowest energy and $\left\langle 100 \right\rangle$ the third lowest in terms of energy. This indicates, $\left\langle 001 \right\rangle$ direction is the easy axis for SCO at all the $U$ values except at $U$=5 eV, being consistent with the experimental observations\cite{2004prl, 2010}.
  
      The difference in total energies of $\left\langle 001 \right\rangle$ and $\left\langle 111 \right\rangle$ directions is found to be significantly high ($\sim$180 meV) at $U$=3 eV, and further reduces with increasing $U$, for example $\sim$118 meV at $U$=4 eV to $\sim$48 meV at $U$=4.4 eV. 
      This energy difference can be considered as a rough estimate of MAE. Such a large value of MAE obatined in the present case ($\sim$48 meV) is consistent with the high coercive field reported in a recent study\cite{2016}. 
      The values of spin quantum number $S$, orbital angular quantum number $L$, total momentum quantum number $J$ and subsequently saturation magnetization and $p_{eff}$ calculated at different $U$ values are provided in Table V for $\left\langle 001 \right\rangle$ direction. Interestingly, changed behavior in variation of values is found at the metal to half-metal (HM) transition ($U$=4.4 eV). With  increase in $U$ from 3 eV, a monotonous increase in values of $S$, $L$, $J$ \& $\mu_z$, $p_{eff}$ is noted till 4.3 eV, with a dramatic decrement in all the values at critical $U$ (4.4 eV) and the decrease in values, continues thereafter.  
$L$ is found to have significantly high values across the $U$ range, varying between 0.21 at $U$=3 eV and a maximum of 0.61 at $U$=4.3 eV.          
The high value of $L$ suggests that the total magnetization of the system cannot be dictated by spin-only moment contribution. Thus, suggesting that the magnetism in SCO cannot be properly understood considering quenched orbital moments.
Notably, the spin MM ($S$) is also affected by application of SOC across the $U$ range, as evident from the Table V, where the $S$ values obtained with SOC are significantly reduced with respect to values obtained without SOC treatment. However, a decrement in $S$ value from $\sim$0.96 (at $U$=4.4 eV) to  $\sim$0.7 is noted above metal to HM phase transition. Here, the calculated $L_z/2S_z$ (0.33) comes out to be notably high with $L_{z}$=0.61 near metal to HM phase transition region. Such high value orbital moment with $L_z/2S_z$=0.25 ($L_{z}$=0.76 $\mu_{B}$) has been previously observed in XMCD study of $\mathrm{SrCo_{0.5}Ir_{0.5}O_{4}}$ containing trivalent Co ions\cite{a}. Therefore, our results put SCO in the family of high orbital MM 3$d$ based transition metal compounds.\cite{a,q,p,c,b,z,r,2019}.  

Considering the prominent role of orbital MM together with spin MM, we determine the magnetization values at each $U$. In order to establish the $U$ and $J$ parameters for this system, the magnetization values and $p_{eff}$ are calculated at the corresponding $U$ values.
 The magnetization value and $p_{eff}$ are calculated as $\sim$2.2 $\mu_B$/f.u and $\sim$3 $\mu_B$, respectively at $U$ = 4.5 eV.   Whereas, highest $p_{eff}$ ($\sim$3.3 $\mu_B$) value is found at $U$=4.3 eV, with total magnetization of $\sim$2.5 $\mu_B$/f.u.
 Considering the experimental results reported on SCO, the studies are found to claim different results on the saturation magnetization value. For instance, the $M(H)$ curve in two earliest of experimental reports on SCO, claim 1.4 $\mu_B$/Co \cite{2005w2} and 1.8 $\mu_B$/Co\cite{2004prl} saturation magnetization values achieved at 5 T and 7 T, respectively. One of the recent studies based on their $M(H)$ curve find the saturation magnetization value less than 1 $\mu_B$/ at 7 T with considerably high $p_{eff}$ (4.26 $\mu_B$)\cite{2013_p2} in contrast to $p_{eff}$ (3.72 $\mu_B$) as reported by $Wang$ $et$ $al$. However, if one examines closely, the $M(H)$ curves in these experimental reports do not seem to have attained saturation at the respective fields. Hence, it would be vague to assume their magnetization values at 5 T or 7 T to be called saturation magnetization values. There appears to be possibility for getting the MM saturated at a higher value, which could be reached at a yet higher magnetic field. If we consider this possibility, then the $\mu_z$ and $p_{eff}$ values obtained around the metal to HM phase transition appear to provide closer description of magnetic properties of this material. This implies $U$($J$) to be around 4.4(1.16) eV for this material, which lies in reasonable range considering the reported Coulomb interaction parameters for 3$d$ transition metal oxides \cite{06,dutta}.\\
Finally, the evolution of FS in SCO is depicted across the correlation induced metal to HM phase transition.
 \begin{figure}[htbp!]
    \centering
    \includegraphics[width=8cm]{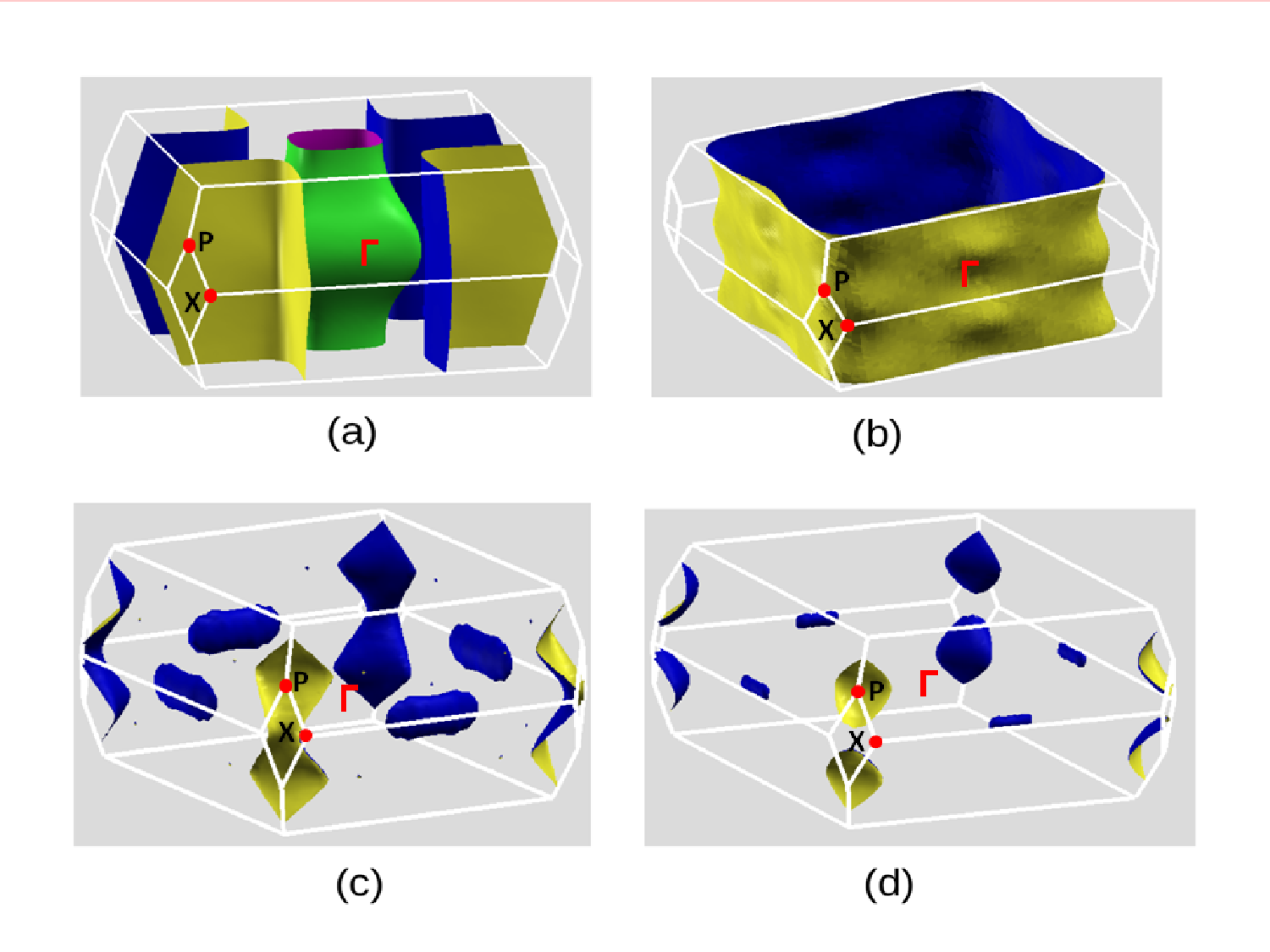}
    \caption{\small FS depiction of S3 solution (a) FS of majority states at $U$=4.3 eV (b) FS of minority states at $U$=4.3 eV, (c) and (d) depict FS of minority states at $U$=4.34 and 4.35 eV, respectively.}

    \label{fig:}
\end{figure}
The changes in FS while approaching the metal to HM phase transition are studied in very close proximity of critical $U$(4.4 eV) (at $U$=4.3, 4.34 \& 4.35 eV) for S3 solution as given in Fig. 4.
The FS consists of the majority and minority sheets, demonstrating its two dimensional characteristics. 
Fig. 4(a) shows the FS of majority states at $U$=4.3 eV where $\Gamma$ centered surface contains electrons and $X$ centered sheets contain holes. The FS of majority states is seen to remains robust across all the considered $U$ values. For this reason we show the FS of majority states at $U$=4.3 eV (in Fig. 4(a)) representative of the FS of majority states at all the other $U$ values $i.e.$ $U$=4.34 \& 4.35 eV. The FS of minority states at the considered $U$ values ($U$=4.3, 4.34 \& 4.35 eV) evolving to HM phase transition state are shown in (Fig. 4(b), 4(c), 4(d)). 
The FS of minority states, shows profound change across the $U$ values, eventually collapsing at critical $U$ (4.4 eV). The FS looks shaped as rectangular cylinder close to Brillouin zone boundary at $U$=4.3 $eV$. While increasing $U$, a substantial collapse in states constituting the FS is noted. For instance, at $U$=4.34 eV, FS contains small sheets at the center of edges and centered around $X$ \& $P$ points. Moreover, significant collapse in states can be observed around $X$ points with a slight increment in $U$ from 4.34 to 4.35 eV.\\
Fig. 4 depicts a gradual collapse in states comprising the FS just within 100 meV window of $U$ values around the critical region. For example, the states around X points collapse at $U$=4.35 eV, whereas states around P points remain robust suggesting the critical $U$ for states around X points to be $U$=4.35 eV, likewise for states around $\Gamma$ points $U$ critical appears to be $U$=4.34 eV.
Generally, the correlation induced metal to insulator phase transitions resulting due to collapse of FS can be either abrupt or gradual. The abrupt collapse of states constituting the Fermi-level suggests the value of $U$ is acting effectively same for all the  \textit{\textbf{k}}-points. On the contrary, the gradual collapse suggests $U$ is acting effectively different for different set of  \textit{\textbf{k}}-points as appears from difference in critical $U$ for different set of \textit{\textbf{k}}-points as visible in this case. This  \textit{\textbf{k}}-point dependent effect of $U$ maybe arising due to significant hybridization effects present in this compound as mentioned before.

\subsection*{Conclusion}

In summary, we succeed in establishing the ground state solution for SCO via a comprehensive mean-field electronic structure study.
Also, the crystal-field splitting, covalency of Co-3$d$ and O-2$p$ orbitals, exchange splitting energy and importance of correlations effects along with spin-orbit coupling are studied in this compound. The DFT calculations give ferromagnetic metallic ground state with MM of 2 $\mu_B$/f.u. Strong covalency of Co-O bond decides the electronic structure properties of SCO. In this work we could strategically use DFT+$U$ to obtain all the possible solution states in SCO to identify its true ground state solution. The DFT+$U$ calculation gives interesting metallic and HM ferromagnetic solution states namely (i) S1 (HM, $S$=1/2), (ii) S2 (Metallic, $S$=1) and (iii) S3 (HM above $U$=4.4 eV, $S$=3/2). 
A characteristic FM state with $S$=3/2 is found to be the ground state solution. The ground state is characterized by a gap of $\sim$0.6 eV appearing in minority Co $d_{xz}$ \& $d_{yz}$ states, showing its HM nature over a range of $U$. 
The compound shows a plausible metal to HM phase transition around a $U$($J$) value of 4.4(1.16) eV. The FS show a continuous gradual collapse in states within small energy window of $\sim$100 meV around $U_c$ suggesting  \textit{\textbf{k}}-dependence of $U_{eff}$.\\
The results suggest interesting physical implications where with a small change in $U$, SCO may be adjusted to function for both metallic and HM phases. This can be achieved through proper doping with suitable correlated atom at Co site.
The HM FM state of SCO in particular can immensely serve due to its potential HM applications in spintronics\cite{spintronics}.
Surprisingly, SCO shows unexpectedly high orbital moment giving $L_z/2S_z$ ratio as high as 0.33 ($L_{z}$=$\sim$0.6) along with large MAE $\sim$48 meV around critical region.
The $U$($J$) value of 4.4(1.16) eV is found to provide realistic description of this compound.
The consideration of orbital moment explains the magnetism in SCO. We find the orbital degrees of freedom along with strong correlation effects playing crucial role in understanding its electronic and magnetic properties. We propose further exploration into orbital degrees of freedom using techniques like XMCD in understanding the role of spin-orbit interaction in this compound.

\centering

\textbf{Supplementary material for \textquotedblleft   Observation of correlation induced metal to half-metal phase transition and large orbital moment in $\mathrm{Sr_2CoO_4}$ \textquotedblright}

\begin{figure}[htbp!]
    \centering
    \includegraphics[width=8cm]{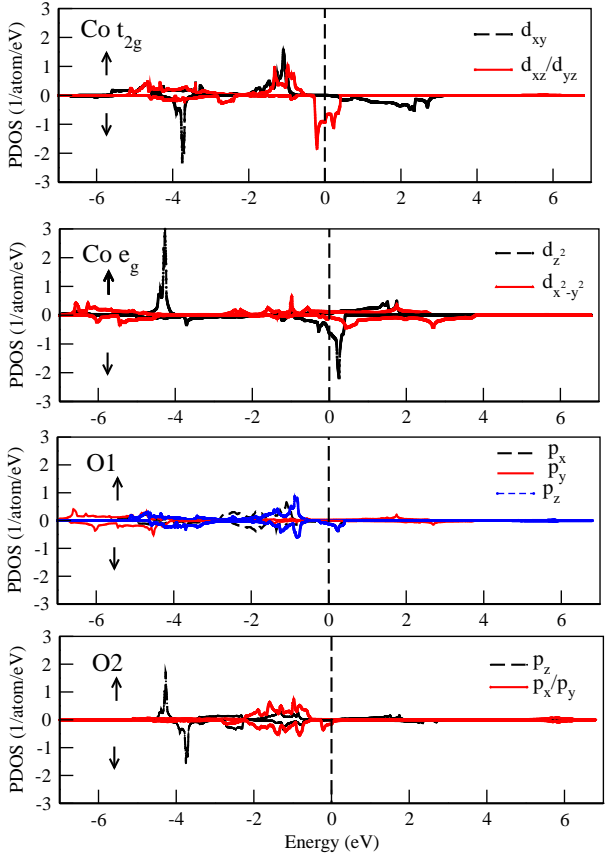}
    \caption{\small DFT calculated PDOS of Co $3d$ and O1 and O2 $2p$, in FM phase of SCO. }

    \label{fig :1 }
\end{figure}

\begin{table*}[htbp!]
    \centering 
    \caption{\small Variation of $J$ (eV) value with $U$ (eV) for SCO.}
  
     \begin{tabular}{|c|c|c|c|c|c|c|c|c|c|}
      \hline
      \tiny $U$ &  \tiny 2   & \tiny 3   & \tiny 4 & \tiny 5 & \tiny 6  \\
     \hline
  \tiny $J$ & \tiny 0.80 & \tiny 0.95  & \tiny 1.08  & \tiny 1.18 & \tiny 1.25 \\

        \hline
     \end{tabular}

    \label{tab:}
\end{table*}

\end{document}